\documentclass[aps, pra, superscriptaddress, twocolumn, amsfonts, amsmath, amssymb, floatfix]{revtex4-2}
\usepackage{graphicx}
\usepackage{float}
\usepackage{sidecap}
\usepackage{dcolumn}
\usepackage{bm}
\usepackage{epsfig}
\usepackage{braket}
\usepackage{color}
\usepackage{ulem}
\usepackage{amsmath}
\usepackage[colorlinks=true, pdfstartview=FitV, linkcolor=blue, citecolor=blue, urlcolor=blue]{hyperref}

\begin{document}

\title{Bayesian-optimized magnetic-field-gradient sensing with a spin-orbit-coupled Bose-Einstein condensate}

\author{Shixiang Chen}
\affiliation{School of Physics, East China Normal University, Shanghai 200241, China}

\author{Juan Wang}
\affiliation{School of Physics, East China Normal University, Shanghai 200241, China}

\author{Yan Li}
\email{yli@phy.ecnu.edu.cn}
\affiliation{School of Physics, East China Normal University, Shanghai 200241, China}
\affiliation{Chongqing Key Laboratory of Precision Optics, Chongqing Institute of East China Normal University, Chongqing 401120, China}

\begin{abstract}

Building on our previous study of the expansion dynamics of a spin-orbit-coupled Bose-Einstein condensate, we propose a scheme for magnetic-field-gradient sensing that relates the maximum rotational deflection angle of the spin-density distribution to the magnetic-field gradient. We analyze the dependence of the sensing performance on key parameters and perform a systematic parameter optimization. A Bayesian optimization algorithm is introduced to efficiently identify the optimal parameter combination, thereby enhancing the sensitivity of magnetic-field-gradient sensing. Numerical simulations under the adopted noise model show that, compared with the parameter set used in our previous work, the optimized gradiometer improves the sensitivity by a factor of five. This work provides a new route for magnetic-field-gradient sensing and offers a reference for machine-learning-assisted quantum precision measurement.

\end{abstract}

\maketitle

%\tableofcontents

\section{Introduction}
  Magnetic field sensing plays an important role in a wide range of fields, including biomedical research, geological resource exploration, and space magnetic-field detection \cite{doi:10.1126/science.172.3990.1329, WOS:000428617600049, bib2,SCHNEIDER201357,WOS:000365729700009}. In recent decades, traditional magnetometers, such as the magnetic needle magnetometer, torsion balance magnetometer, and Hall magnetometer, can no longer meet the ever-increasing demands for precision. Researchers have gradually turned their attention to quantum magnetometers, which offer high sensitivity and precise control over quantum states. Representative advances include superconducting quantum interference device (SQUID) magnetometers based on the Josephson effect \cite{PhysRevLett.12.159, PhysRevLett.111.067202, PhysRevLett.123.143003, WOS:000312488200050, 10.1063/5.0245555}, optical pumping magnetometers utilizing Zeeman level splitting \cite{PhysRev.107.1559, PhysRevLett.6.280, doi:10.1126/sciadv.aba8792, PhysRevLett.120.033202}, coherent population trapping (CPT) magnetometers employing atomic interference \cite{PhysRevLett.69.1360, A.Nagel_1998, PhysRevLett.105.193601, Maldonado:26}, nonlinear magneto-optical rotation (NMOR) magnetometers based on the interaction of light and magnetic fields \cite{GAWLIK1974283, PhysRevA.73.053404, PhysRevApplied.12.011004, Tian:23, doi:10.1126/sciadv.1700422}, and spin-exchange relaxation-free (SERF) magnetometers \cite{PhysRevLett.121.091802, PhysRevA.111.032602}.

However, many applications require not only the measurement of magnetic-field strength but also, more importantly, the precise acquisition of magnetic-field gradients. Compared with magnetic field measurements, magnetic-field-gradient measurements can provide higher spatial resolution \cite{PhysRevApplied.21.014003, PhysRevLett.117.138501}. For example, optically pumped magnetometers (OPMs) have been employed to achieve high-sensitivity, high-resolution detection in magnetoencephalography \cite{doi:10.1126/sciadv.aba8792} and magnetocardiography \cite{10.1063/1.5091007}. In cold-atom physics, magnetic-field and magnetic-field-gradient sensing schemes based on OPM, CPT, and NMOR have been extensively investigated \cite{article12,doi:10.1126/sciadv.adt3938,Yin:26}. These schemes generally extract magnetic-field information from internal-state populations, atomic phases, or resonance-frequency shifts. 

Compared with hot-atom magnetometers, cold-atom sensors generally have lower absolute sensitivity and require more complex experimental apparatus and costly preparation procedures \cite{article12,PhysRevApplied.22.024047, PhysRevLett.104.093602}. Nevertheless, they offer high spatial resolution and precise control over atomic quantum states, making them promising platforms for spatially resolved magnetic-field and magnetic-field-gradient sensing \cite{doi:10.1126/sciadv.adt3938, Wildermuth2005}. In particular, spin-orbit coupled (SOC) Bose-Einstein condensate (BEC) provides a highly controllable platform in which the coupling between atomic spin and motion enables the engineering of rich quantum states and dynamics \cite{WOS:000287924100038}.

Our previous work showed that the expansion of an SOC BEC under a position-dependent detuning produces a measurable deflection of the spin-density distribution \cite{PhysRevA.110.043307}. The detuning gradient is related to the magnetic-field gradient, providing a possible method for magnetic-field-gradient detection. In that scheme, the gradient response can be extracted from the spin-density distribution using time-of-flight (TOF) imaging \cite{science.275.5300.637}, enabling a relatively direct geometric readout. However, the sensitivity depends jointly on the trapping frequency  $\omega_0$, the detuning gradient $\eta$ and the Raman coupling strength $\Omega $. Efficient multiparameter optimization is therefore required.

Machine learning provides an effective tool for addressing such complex multiparameter optimization problems. In recent years, machine-learning methods have been applied to parameter optimization, autonomous experimental control, BEC preparation, and quantum-state generation in cold-atom systems \cite{Li2022, Reinschmidt2024, PhysRevX.8.031086, PhysRevResearch.4.043216, Liang2018}. Among these methods, Bayesian optimization is particularly suitable for problems in which numerical simulations or experimental evaluations are computationally expensive. It constructs a probabilistic surrogate model of the objective function from previously acquired data and adaptively selects subsequent evaluation points, thereby enabling efficient exploration of complex parameter spaces with a limited number of evaluations \cite{article12, PhysRevApplied.22.044058}.

    In this work, we propose a magnetic-field-gradient sensing scheme based on the expansion dynamics of a two-component SOC BEC and employ Bayesian optimization to enhance the sensitivity of the gradiometer. Using the parameter set adopted in our previous work \cite{PhysRevA.110.043307}as a reference, the sensitivity evaluated with the noise model introduced here is $4.65 \times 10^{-6} \, \mathrm{T/cm\cdot \sqrt{Hz}}$. Bayesian optimization improves the sensitivity to $9.30 \times 10^{-7} \, \mathrm{T/cm\cdot \sqrt{Hz}}$, corresponding to an improvement by a factor of approximately five.

	This paper is organized as follows. Section II introduces the theoretical model of our SOC system with zero-momentum phase. Section III presents the parameter dependence of magnetic-field-gradient sensing and the optimization results. Section IV concludes the paper.

\section{Theoretical model}

We adopt the scheme developed by the Spielman group \cite{WOS:000287924100038} to investigate the SOC BEC. Specifically, we confine $^{87}$Rb atoms in a harmonic trap potential given by $V_{trap} = \frac{1}{2} m \omega_x^2 x^2 + \frac{1}{2} m \omega_y^2 y^2 + \frac{1}{2} m \omega_z^2 z^2$ , where  $m$  is the atomic mass of $^{87}$Rb, and  $\omega_x$,  $\omega_y$, and  $\omega_z$  are the corresponding trapping frequencies along the x- , y- , and  z-directions, respectively. To ensure the system exhibits quasi-two-dimensional dynamics, we set $\omega_z \gg \omega_x, \omega_y$. The single-particle Hamiltonian in the presence of a synthetic gauge field is \cite{PhysRevA.79.063613,Lin2009}:

	\begin{equation}
		\hat{H}_{\text{sp}} = \frac{1}{2m} \left( \hat{p}_x - \hbar k_r \sigma_z \right)^2 + \frac{\hat{p}_y^2}{2m} + V_{\text{trap}} - \frac{\Omega}{2} \sigma_x - \delta(y) \sigma_z,\label{eq1}
	\end{equation}
where $\hat{p}_x$  and  $\hat{p}_y$  are the canonical momentum operators along the  x- and  y-directions, respectively. The $\hbar k_r$ denotes the photon recoil momentum, and  $\Omega$  represents the Raman coupling strength.  $\delta(y)$  is the two-photon detuning, which depends on the y -coordinate.

In this work, we focus on the zero-momentum phase of the SOC BEC, which occurs under the condition $\Omega > \Omega_c \equiv 4E_r$ , where  $E_r = \hbar^2 k_r^2 / 2m$  is the recoil energy. For a sufficiently small detuning gradient $\eta$ and an appropriate Raman coupling strength, vortices do not emerge, and the total density of the condensate can be approximately described by a Thomas-Fermi (TF) distribution. Under this condition, the quantum pressure term can be neglected when applying the hydrodynamic theory. Consequently, the superfluid velocity is given by \cite{PhysRevLett.120.183202, PhysRevA.108.053316}:

	\begin{equation}
		v=\left(\frac{\hbar}{m^\ast}\nabla_x\phi-\frac{\eta}{\hbar}\frac{\Omega_c}{\Omega}\ y,\frac{\hbar}{m}\nabla_y\phi,\ \frac{\hbar}{m}\nabla_z\phi\right).\label{eq2}
	\end{equation}

Here, \( m^\ast = m\left(1 - \Omega_c / \Omega\right)^{-1} \) is the effective mass in the zero-momentum phase, and  $\phi$  is the phase, satisfying the condition  $\sqrt{m / m^\ast} \, \omega_x = \omega_y \equiv \omega_0$. The detuning-related term  -$\frac{\eta}{\hbar} \frac{\Omega_c}{\Omega} y$  is the key factor responsible for generating the rotational velocity field. When the detuning gradient  $\eta$  is non-zero, this term—which is proportional to the y -coordinate—directly induces a rotational velocity field in the $ x\text{-}y$  plane.

    \begin{figure}
		\centering
		\includegraphics[width=1\columnwidth]{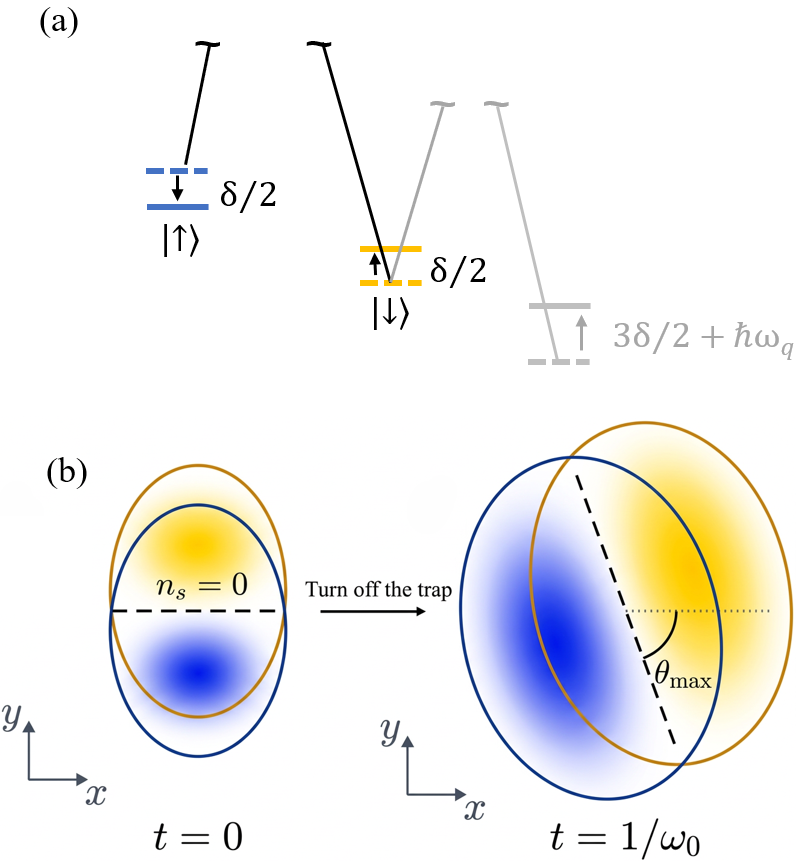}
		\caption{\rmfamily (a) Level diagram for Raman coupling between the two pseudospin states
$|\uparrow\rangle$ and $|\downarrow\rangle$ of SOC BEC, with two-photon detuning
$\delta$. The quadratic Zeeman shift \(\hbar\omega_q\) detunes the third Zeeman state from Raman resonance, allowing it to be neglected in the effective two-level description.
(b) Rotational expansion of the two-component SOC BEC.
Here, $n_s=n_1-n_2$ denotes the spin density, where $n_1$ and $n_2$
are the densities of the two components. The black dashed line indicates
the $n_s=0$ boundary. After the in-plane trapping potential is switched
off, the rotation of this boundary at $t=1/\omega_0$ relative to its
initial direction defines the maximum deflection angle $\theta_{\max}$}.
		\label{fig1}
	\end{figure}
	
     %This results in a time-dependent scattering length: \(a(t) = a_{bg} \left( 1 - \frac{\Delta}{\bar{B} + \delta B \cos(\omega_m t) - B_0} \right)\). By performing a first-order Taylor expansion, we obtain:}
    %\begin{equation}
     %a(t)=a_{bg}\left ( 1-\frac{\Delta}{\bar{B}-B_0 }  \right ) +\frac{\Delta \cdot \delta B}{B_{1}^{2}} a_{bg} \cos (\omega_m t).
	%\end{equation}

     At time $t = 0$, we switch off the confinement potentials along the $x$ and $y$ directions, allowing the two-component BEC to undergo free expansion under the velocity field. As shown in \textcolor{blue}{Fig.}\ref{fig1}\textcolor{blue}{(b)}, we find that the spin-density deflection angle $\theta$ reaches its maximum value at the system time $t = 1/\omega_0$ \cite{PhysRevA.110.043307}:

	\begin{equation}
		\theta_{\mathrm{max}}=\frac{\pi}{2}-\arctan{\left(\eta/\hbar\omega_0\left(2-\Omega_c/\Omega\right)\right)}.\label{eq3}
	\end{equation}

The maximum value of this deflection angle depends on multiple parameters. Given the relationship between the magnetic field gradient and the detuning: 

	\begin{equation}
		\nabla B=-\eta k_r/\left(g_F\mu_B\right), \label{eq4}
	\end{equation}
where $g_F$ is the Land\'e $g$-factor of the relevant hyperfine state and
$\mu_B$ is the Bohr magneton. Our system therefore indirectly measures the magnitude of the magnetic field gradient by determining the maximum deflection angle. The dynamics of the SOC BEC rely on the time-dependent Gross-Pitaevskii (GP) equations:

\begin{subequations}\label{eq:GP}
\begin{align}
\begin{aligned}
i\hbar\frac{\partial\psi_1}{\partial t}
&= \bigg[
-\frac{\hbar^2(k_x-k_r)^2}{2m}
+\frac{p_y^2}{2m}
+{}\\
&\quad g\left(|\psi_1|^2+|\psi_2|^2\right)
\bigg]\psi_1
+\frac{\hbar\Omega}{2}\psi_2
-\eta k_r y\psi_1,
\end{aligned}
\label{eq:GP-a}
\\[1ex]
\begin{aligned}
i\hbar\frac{\partial\psi_2}{\partial t}
&= \bigg[
-\frac{\hbar^2(k_x+k_r)^2}{2m}
+\frac{p_y^2}{2m}
+{}\\
&\quad g\left(|\psi_1|^2+|\psi_2|^2\right)
\bigg]\psi_2
+\frac{\hbar\Omega}{2}\psi_1
+\eta k_r y\psi_2.
\end{aligned}
\label{eq:GP-b}
\end{align}
\end{subequations}
    
The mean-field interaction energy of the condensate is characterized by$ V_{\mathrm{int}}=\frac{1}{2}\int d\mathbf{r}\sum_{\alpha,\beta}{g_{\alpha\beta}n_\alpha n_\beta}$, where $n_{\alpha,\beta}$ is the particle density of each component and $g_{\alpha\beta}=4\pi\hbar^2 a_{\alpha\beta}/m$ are the interaction constants in different spin channels with $a_{\alpha\beta}$ the corresponding $s-$wave scattering lengths. In this work, we assume isotropic coupling constants $g_{\alpha\beta}\equiv g$. We obtain the sensitivity for detecting the magnetic-field gradient by numerically solving the Gross–Pitaevskii (GP) equations.

\section{Results}
\subsection{Sensitivity analysis}
	\label{subsec1}
The sensitivity $S_{\nabla B}$ of the gradient magnetometer is determined by three parameters: the response $\mathcal{R}$ of the system to the magnetic field gradient, the uncertainty $\sigma_\theta$ of the deflection angle, and the single-cycle time $T_c$, which reflects the repetition rate of measurements. Their relationship is given by \cite{RevModPhys.92.021001}:
	\begin{equation}
		S_{\nabla B}=\mathcal{R}\cdot\sigma_\theta\cdot\sqrt{T_c} \label{eq6}
	\end{equation}

      \begin{figure}
		\centering
		\includegraphics[width=1\columnwidth]{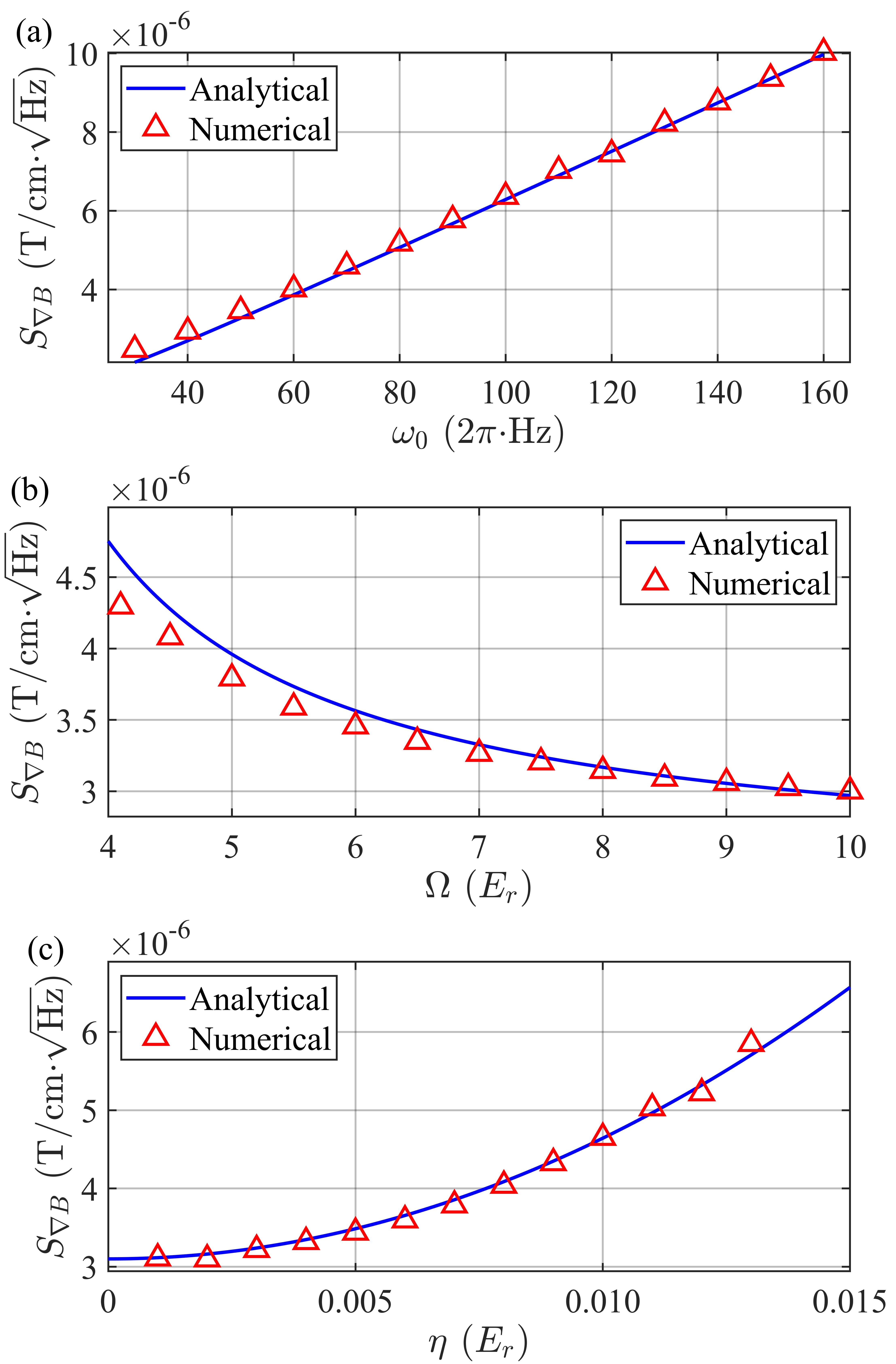}
		\caption{\fontfamily{ptm}\selectfont Dependence of the sensitivity $S_{\nabla B}$ on individual parameters while keeping the others fixed. (a) Sensitivity as a function of the trapping frequency  $\omega_0$ , for fixed  $\Omega = 6E_r$  and  $\eta = 0.005E_r$ . (b) Sensitivity as a function of  $\Omega$ , for fixed  $\eta = 0.005E_r$  and  $\omega_0 = 2\pi \times 50 \, \text{Hz}$ . (c) Sensitivity as a function of $\eta$, for fixed  $\Omega = 6E_r$  and  $\omega_0 = 2\pi \times 50 \, \text{Hz}$.}
		\label{fig2}
	\end{figure}

The cycle time $T_c$ consists of the BEC preparation time $T_{\mathrm{prep}}$, the expansion evolution time $T_{\mathrm{evo}} = 1/\omega_0$, and the imaging readout time $T_{\mathrm{det}}$. The angular uncertainty $\sigma_\theta$ originates from two noise contributions in the sensing process. The first type is the particle number shot noise: the detected atom number of each spin component at every grid point is sampled from a Poisson distribution $n_{\alpha\beta}^{\mathrm{obs}} \sim \mathrm{Poisson}\bigl( n_{\alpha\beta}^{clean} \bigr)$ \cite{Vibel_2024, Kristensen_2017}, where $n_{\alpha,\beta}^{\mathrm{clean}}$ denotes the noise-free density distribution of each component. The second type is camera readout noise. It is modeled as independent Gaussian noise: $n_{\alpha\beta}^{\mathrm{img}} \sim \mathcal{N}\left(0, \sigma_{\mathrm{img}}^2\right)$, where $\sigma_{\mathrm{img}}$ characterizes the readout-noise level of the imaging system \cite{Vibel_2024, PhysRevA.82.061606}. To quantify the combined effect of these two types of noise on the angular measurement, we employ a Monte Carlo method: at the optimal expansion time $t = 1/\omega_0$, we take the noise-free spin density distribution $n_{\alpha\beta}^{\mathrm{clean}}$, independently apply Poisson sampling to each grid point to simulate the particle number noise $n_{\alpha\beta}^{\mathrm{obs}}$, and then add Gaussian noise to simulate the imaging noise $n_{\alpha\beta}^{\mathrm{img}}$. The total density distribution of the system is then given by $n_{\alpha\beta}^{\mathrm{final}} = n_{\alpha\beta}^{\mathrm{obs}} + n_{\alpha\beta}^{\mathrm{img}}.$
The above Monte Carlo simulation procedure is independently repeated $N_{\mathrm{MC}} = 1000$ times, and the standard deviation of the resulting $\theta_{\max}$ values gives the angular measurement uncertainty $\sigma_\theta$.

From \textcolor{blue}{Eqs.} \ref{eq3} and \ref{eq4}, we obtain the expression for the response $\mathcal{R}$ of the magnetic field gradient with respect to the deflection angle:

\begin{equation}
\mathcal{R}
\equiv
\left|
\frac{\partial \nabla B}{\partial \theta_{\mathrm{max}}}
\right|
=
\left|
\frac{\hbar\omega_0 k_r}{g_F\mu_B\Delta}
\left[
1+
\left(
\frac{\eta\Delta}{\hbar\omega_0}
\right)^2
\right]
\right|,
\label{eq7}
\end{equation}
here, $\Delta \equiv  2 - \frac{\Omega_c}{\Omega}$. From \textcolor{blue}{Eqs.} \ref{eq6} and \ref{eq7}, the sensitivity $S_{\nabla B}$ of the gradient magnetometer is influenced by three tunable parameters: the Raman coupling strength $\Omega$, the trap frequency $\omega_0$, and the detuning $\eta$. To characterize their individual effects, we vary one parameter at a time while keeping the other two fixed.

As shown in \textcolor{blue}{Fig.}\ref{fig2}\textcolor{blue}{(a)}, with $\Omega = 6E_r$ and $\eta = 0.005E_r$ fixed, $\omega_0$ is varied to observe the change in sensitivity. The analytical result demonstrates that $S_{\nabla B}$ is linearly proportional to $\omega_0$, in agreement with the numerical results.

\textcolor{blue}{Figure} \ref{fig2}\textcolor{blue}{(b)} presents the sensitivity as a function of the Raman coupling strength $\Omega$ when $\eta = 0.005E_r$ and $\omega_0 = 2\pi \times 50\,\mathrm{Hz}$. In that case, $S_{\nabla B}$ exhibits a nonlinear negative correlation with $\Omega$. When $\Omega$ lies in the low-frequency region, a discrepancy of up to 12\% emerges between the analytical and numerical results. This is attributed to $\Omega$ approaching the minimum value of the zero-momentum phase, $4E_r$, where the expansion velocities along the x and y directions differ significantly, making the system susceptible to vortex formation.

In \textcolor{blue}{Fig.}\ref{fig2}\textcolor{blue}{(c)}, with $\Omega = 6E_r$ and $\omega_0 = 2\pi \times 50\,\mathrm{Hz}$ fixed, $\eta$ is varied to observe the change in sensitivity. The analytical result,
$\frac{\partial\nabla B}{\partial\theta_{\mathrm{max}}} = \frac{\hbar\omega_0 k_r}{g\mu_B} \cdot \frac{1}{\Delta} \left[ 1 + \left( \frac{\xi \eta \Delta}{\hbar\omega_0} \right)^2 \right]$
agrees well with the numerical data, here, $\xi=1.5$ is a correction factor. In the high-$\eta$ regime, the velocity difference between the x and y directions becomes substantial, which also favors vortex generation. In this regime, the system's sensitivity no longer fully adheres to the description of \textcolor{blue}{Eq.} \ref{eq7}, necessitating this correction.

      \begin{figure}
		\centering
		\includegraphics[width=1\columnwidth]{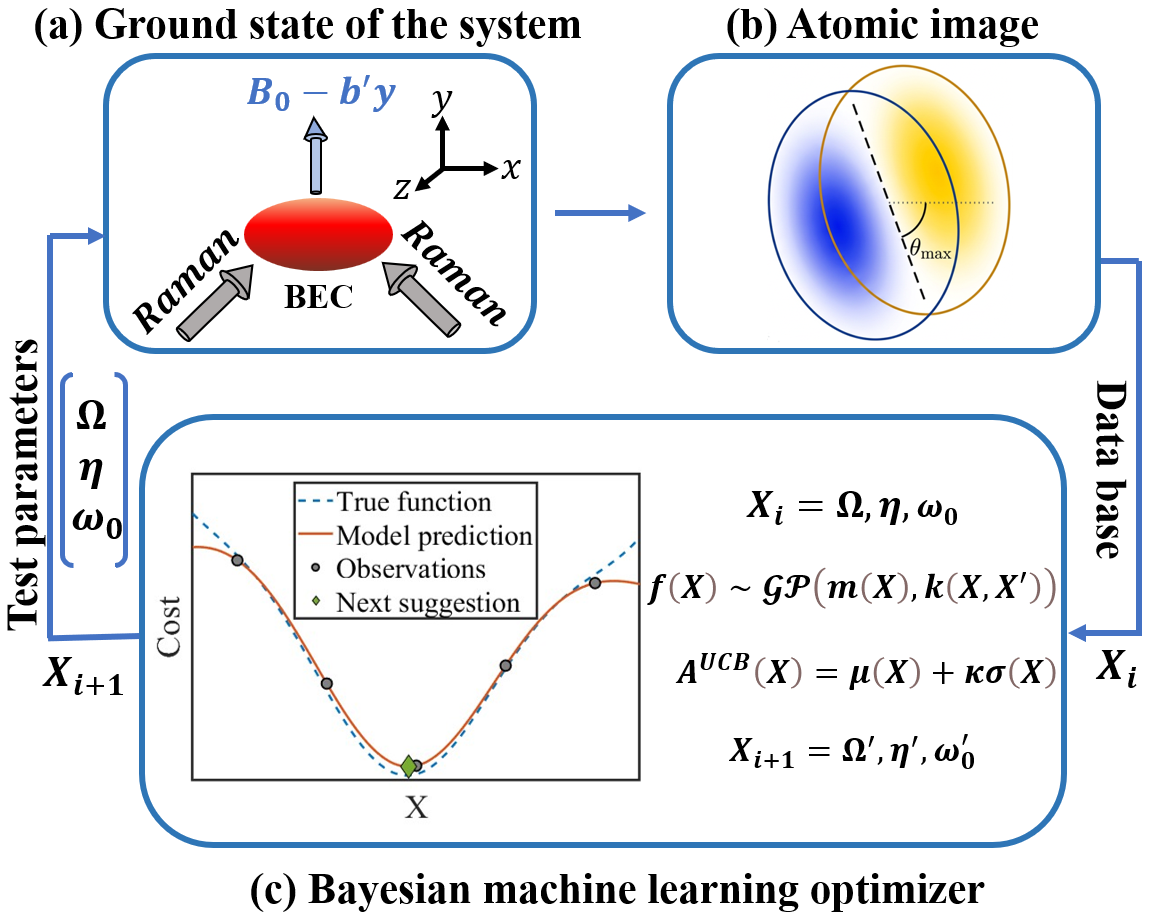}
		\caption{\fontfamily{ptm}\selectfont Diagram of a Bayesian-algorithm-based parameter optimization loop. (a) Ground state of the SOC BEC. (b) Spin‑density distribution of the expanding and rotating two‑component BEC after the trapping potential is removed. (c) Bayesian machine learning optimizer.}
		\label{fig3}
	\end{figure}

    \subsection{Bayesian optimizer}
	\label{subsec2}

It should be noted that the analytical sensitivity results discussed above are not universal; they agree well with the numerical results only within an appropriate range and under the condition of varying a single parameter. Crucially, if the two fixed parameters are not appropriately chosen, the discrepancy between \textcolor{blue}{Eqs.} \ref{eq6} and the numerical results may increase. For instance, if the fixed parameters in \textcolor{blue}{Fig.}\ref{fig2}\textcolor{blue}{(a)} are changed from  $\Omega = 6E_r$, $\eta = 0.005E_r$  to  $\Omega = 6E_r$, $\eta = 0.010E_r$ , the relationship in the low-frequency region of  $\omega_0$  would no longer be linear but instead exhibit a nonlinear abrupt change. Therefore, \textcolor{blue}{Eq.} \ref{eq7} is not universally applicable across the entire three-parameter space. Moreover, when attempting to derive the two- or three-parameter equations based on \textcolor{blue}{Eq.} \ref{eq7}, solving for the extrema of multivariate functions becomes extremely difficult.

Consequently, we employ a Bayesian algorithm to construct a feedback loop within the multidimensional parameter space, thereby obtaining the optimal sensitivity value.

        	\begin{figure}
		\centering
		\includegraphics[width=1\columnwidth]{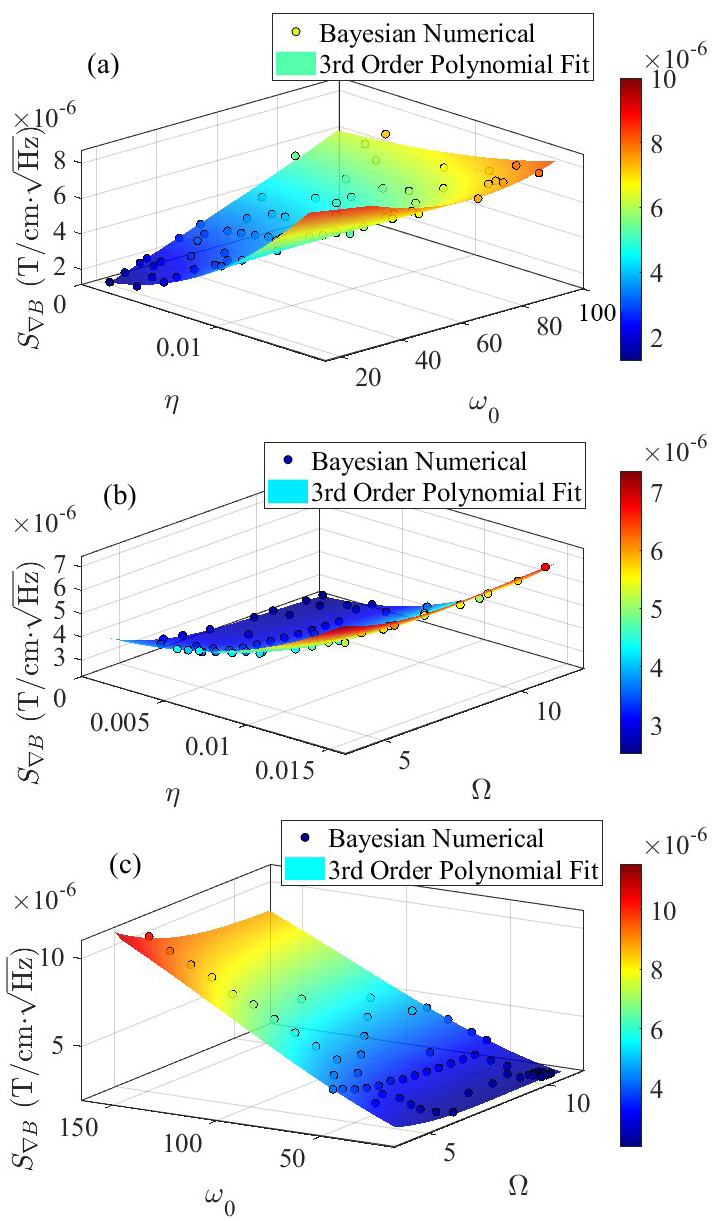}
		\caption{\fontfamily{ptm}\selectfont Dependence of the sensitivity on the two-parameter combinations $\Omega,\eta$ and $\omega_0, \eta$, along with the corresponding fitted surfaces. (a) Sensitivity as a function of the trapping frequency $\omega_0$ and the detuning $\eta$, with $\Omega$ fixed at $6E_r$. (b) Sensitivity as a function of the detuning $\eta$ and the Raman coupling strength $\Omega$, with $\omega_0$ fixed at $2\pi \times 50$ Hz. (c) Sensitivity as a function of the trapping frequency $\omega_0$ and the Raman coupling strength $\Omega$, with $\eta$ fixed at 0.005 $E_r$.}
		\label{fig4}
	\end{figure}

%\textcolor{blue}{Figure} \ref{fig3} illustrates the Bayesian optimization workflow that we adopt in this work. The objective of this optimization problem is to minimize the sensitivity \(S_{\nabla B}(X)\) of the gradient magnetometer, where the parameter set is given by \(X = (\Omega, \eta, \omega_0)\). Bayesian optimization constructs a probabilistic surrogate model to approximate the unknown sensitivity function and leverages this model to guide the search efficiently.

%The specific procedure is as follows. First, a set of initial parameter points \(X_i\) is evaluated by numerically solving the Gross–Pitaevskii equation to obtain the corresponding sensitivity values \(S_{\nabla B}(X_i)\), which constitute the initial training dataset. A Gaussian process model is then fitted to these data, providing a probabilistic description of the sensitivity function across the parameter space. Next, numerical optimization is performed on the model's predictions to determine the next most promising parameter combination \(X_{i+1}\). This predicted optimum is then tested by solving the numerical GP equation, yielding a new sensitivity observation \(S_{\nabla B}(X_{i+1})\). This observation is added to the training set, and the model is refitted to initiate the next iteration. This closed loop of "evaluation → modeling → optimization → evaluation" enables the system to rapidly approach the globally optimal parameter region with as few numerical simulations as possible, thereby improving the sensitivity of the gradient magnetometer.

\textcolor{blue}{Figure} \ref{fig3} illustrates the Bayesian optimization workflow used in this work. We formulate the optimization as the minimization of the gradient sensitivity $S_{\nabla B}(\mathbf{X})$ over the parameter set $\mathbf{X}=(\Omega,\eta,\omega_0)$. The optimizer is initialized with a historical dataset of previously evaluated parameter combinations and their corresponding sensitivities. Each sensitivity value is obtained by numerically solving the GP equation.

A Gaussian-process surrogate is fitted to the available data to
model the dependence of the transformed objective on the control
parameters. Since the Bayesian optimization routine maximizes its objective, the sensitivity values are multiplied by \(-1\), so that maximizing the transformed objective is equivalent to minimizing the original sensitivity $S_{\nabla B}$. After $i$ evaluations, the Gaussian-process provides a
posterior mean $\mu_i(\mathbf{X})$ and posterior standard deviation
$\sigma_i(\mathbf{X})$ for each candidate parameter set.

The optimizer uses the upper confidence bound (UCB) acquisition
function \cite{6138914}: $A_i^{\mathrm{UCB}}(\mathbf{X})=\mu_i(\mathbf{X})+\kappa\sigma_i(\mathbf{X}),$
where $\kappa=2.576$ controls the balance between exploiting candidates with
high predicted objective values and exploring candidates with high
predictive uncertainty.     The next candidate is selected by maximizing the UCB acquisition function: $\mathbf{X}_{i+1}
=
\underset{\mathbf{X}\in\mathcal{D}}{\operatorname{arg\,max}}
\left[A_i^{\mathrm{UCB}}(\mathbf{X})\right],$
where \(\mathcal D\) denotes the bounded parameter space explored by the optimizer. In this work, \(\Omega/E_r\in[4,12]\), \(\eta/E_r\in[0.001,0.020]\), and \(\omega_0/(2\pi)\in[15,160]\ \mathrm{Hz}\).

The selected parameter combination is then evaluated by numerically solving the GP equations. The resulting sensitivity is added to the historical dataset, and the Gaussian-process surrogate is updated. After that, the selection and evaluation steps are repeated.

           	\begin{figure}
		\centering
		\includegraphics[width=0.95\columnwidth]{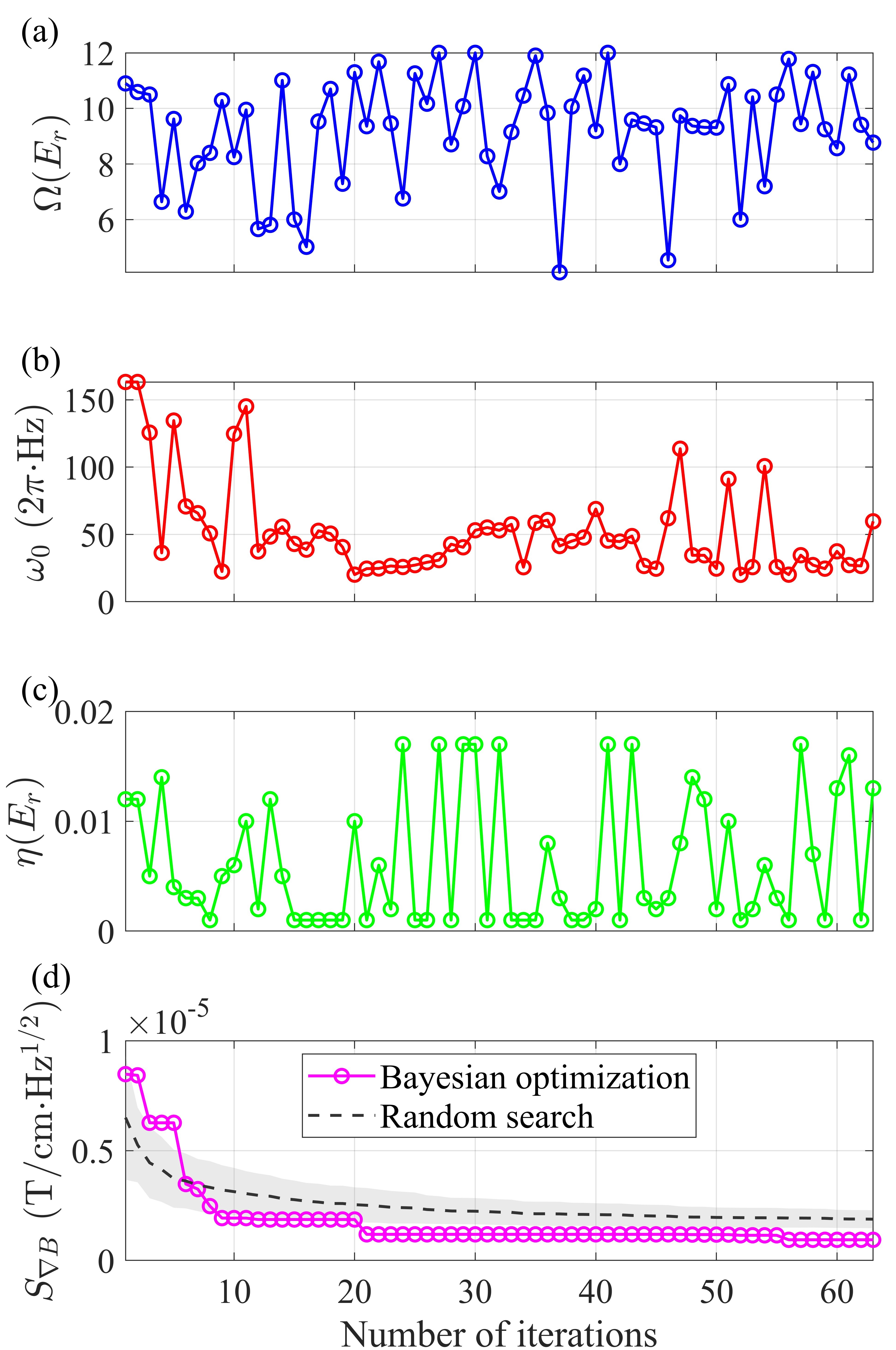}
		\caption{\fontfamily{ptm}\selectfont Three-parameter Bayesian optimization process. (a) Evolution of the Raman coupling strength \(\Omega\) with the number of iterations; (b) evolution of the trapping frequency \(\omega_0\) with the number of iterations; (c) evolution of the detuning gradient \(\eta\) with the number of iterations. (d) Convergence of the magnetic-field-gradient sensitivity \(S_{\nabla B}\) as a function of the iteration number. The purple solid line shows the cumulative minimum sensitivity obtained during the Bayesian optimization process. The black dashed line and gray shaded region represent the random-search baseline simulated using a Gaussian-process surrogate model \cite{Wang2025Gaussian}. Specifically, the dashed line shows the mean best-so-far sensitivity obtained from 100 independent random-search trials, while the shaded region indicates the mean \(\pm 1\sigma\), where \(\sigma\) is the standard deviation across the 100 trials.}
		\label{fig5}
	\end{figure}

  The two-parameter optimization results are presented in \textcolor{blue}{Fig.}\ref{fig4}. The feedback loop and cross-training capabilities of the Bayesian algorithm enable the identification of optimal parameter choices with a relatively small dataset. Using the sensitivity data obtained from the optimization, we construct a two-parameter model for each pair of parameters. The relationship between the two parameters and sensitivity was visualized by fitting a third-order polynomial surface. With $\Omega$ fixed at 6$E_r$, the optimal sensitivity obtained by varying $\omega_0$ and $\eta$ is $S_{\nabla B} = 2.12\times10^{-6}$ T/cm $\cdot$ $\sqrt{Hz}$, corresponding to a 54\% improvement in sensitivity.

Through the three-parameter Bayesian optimization training, the system leverages a feedback loop to efficiently approach the global optimum with a relatively small number of evaluations. As shown in \textcolor{blue}{Fig.}\ref{fig5}, the purple solid line depicts the convergence of the cumulative minimum sensitivity obtained during the Bayesian optimization process. As the number of iterations increases, the Bayesian optimizer adaptively adjusts the parameter combination of the Raman coupling strength \(\Omega\), the trapping frequency \(\omega_0\), and the detuning \(\eta\) based on the sensitivity results from each iteration. After a number of iterations, the sensitivity of the gradient magnetometer rapidly converges to \(S_{\nabla B} = 9.3 \times 10^{-7} \, \mathrm{T/cm\cdot \sqrt{Hz}}\).

For comparison, the black dashed line and shaded region represent the random search baseline simulated using a Gaussian process surrogate model (mean \(\pm 1\sigma\) over 100 independent trials). At the early stage of the optimization, the random search occasionally provides a sensitivity lower than the current value obtained by Bayesian optimization, owing to its independent random sampling from a candidate pool containing relatively favorable parameter predictions. However, as the number of iterations increases, the convergence speed of the random search becomes considerably slower than that of Bayesian optimization due to the absence of a feedback mechanism. After the same number of iterations, the average best sensitivity achieved by the random search is only approximately \(2.5 \times 10^{-6} \, \text{T/cm} \cdot \sqrt{\text{Hz}}\), which is about 2.7 times that of the final sensitivity obtained by Bayesian optimization. This comparison fully validates the sample efficiency advantage of Bayesian optimization: it is capable of identifying a better sensitivity with significantly fewer evaluations.

At low trapping frequencies and large position-dependent detuning gradients, the two spin components exhibit strongly inhomogeneous expansion and velocity fields. In this regime, the Thomas-Fermi approximation no longer provides an accurate description of the dynamics. And the emergence of vortex leads to an abrupt change in the system's sensitivity to magnetic field measurement, causing it to degrade from the order of $ 10^{-6}\,\mathrm{T/cm\cdot \sqrt{Hz}}$ to $10^{-4}\,\mathrm{T/cm\cdot \sqrt{Hz}}$. Therefore, it is necessary to identify the high-sensitivity dynamic range in which vortex does not form.

\begin{figure}
		\centering
		\includegraphics[width=0.95\columnwidth]{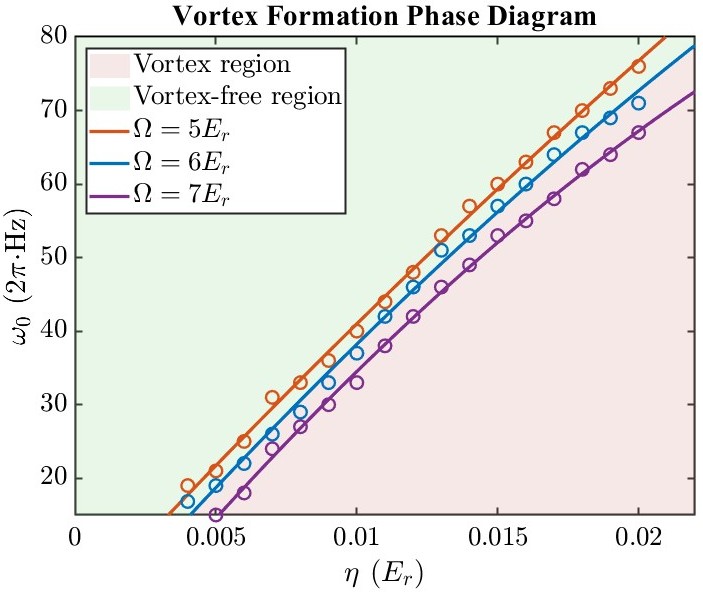}
		\caption{\fontfamily{ptm}\selectfont Vortex-nucleation boundaries in the parameter space of the
detuning gradient $\eta$ and the trapping frequency $\omega_0$ for
different Raman coupling strengths. The orange, blue, and purple
curves correspond to $\Omega=5E_r$, $6E_r$, and $7E_r$,
respectively. The shaded regions are determined by the fitted
vortex-nucleation boundary at $\Omega=6E_r$.}
		\label{fig6}
	\end{figure}

As shown in \textcolor{blue}{Fig.}\ref{fig6}, we map the vortex formation phase diagram of the system by selecting parameter combinations at the sensitivity mutation points. Taking the two parameters, trapping frequency $\omega_0$ and detuning $\eta$, as an example, the system is more susceptible to vortex formation at lower $\omega_0$ and higher $\eta$. For a given $\eta$, we select the point with the maximum $\omega_0$ from the parameter sets that still generate vortices as a boundary point. A series of such boundary points are fitted with a second-order polynomial to define the vortex phase boundary (blue curve in the figure). The fitting equation is $\omega_0 = -30108\eta^2 + 4350\eta - 2.3$. When the chosen $\omega_0$ for a selected $\eta$ falls below the value predicted by this fitting equation, vortex appears in the system.

\section{Conclusions}

We have investigated the roles of key parameters in a magnetic-field-gradient sensing scheme based on the rotational expansion dynamics of SOC BEC. The results show that the sensitivity depends jointly on the Raman coupling strength $\Omega$, the trapping frequency $\omega_0$, and the detuning gradient $\eta$. Although the analytical response relation describes the numerical results well within a limited parameter range, deviations between the numerical results and the analytical predictions arise near the boundary of the zero-momentum phase and in the regime of low trapping frequency and large detuning. These observations indicate that the analytical expression for the sensitivity cannot predict the sensor performance across the full parameter space and is valid only within specific parameter regimes. More effective optimization approaches are therefore required to determine the relevant parameters.

Using Bayesian optimization, we search the three-dimensional parameter space to minimize the magnetic-field-gradient sensitivity. For the parameter set used in our previous work, the sensitivity evaluated with the noise model adopted here is $4.65\times10^{-6}~\,\mathrm{T/cm\cdot \sqrt{Hz}}$. With $\Omega=6E_r$ fixed, two-parameter optimization over $\omega_0$ and $\eta$ yields a sensitivity of $2.12\times10^{-6}~\,\mathrm{T/cm\cdot \sqrt{Hz}}$. Optimizing all three parameters further improves the sensitivity to $9.3\times10^{-7}~\mathrm{T/cm\cdot \sqrt{Hz}}$, approximately five times better than that obtained with the previous parameter set. For the same number of evaluations, Bayesian optimization also converges faster than random search, demonstrating its higher sample efficiency.

We further identify vortex formation as a key factor limiting the operational parameter range of the sensor. In the regime of low $\omega_0$ and high $\eta$ regime, vortex formation substantially degrades the sensitivity from the order of $10^{-6}\,\mathrm{T/cm\cdot \sqrt{Hz}}$ to that of $10^{-4}\,\mathrm{T/cm\cdot \sqrt{Hz}}$. For $\Omega=6E_r$, the fitted vortex boundary shown in \textcolor{blue}{Fig.}\ref{fig6} delineates the parameter region where the vortex-free description remains valid. This boundary provides a physical constraint on parameter selection and defines the effective operating range in which the gradient magnetometer retains high sensitivity.

Our results demonstrate that the rotational deflection dynamics during the expansion of an SOC BEC can be exploited for magnetic-field-gradient sensing, and that Bayesian optimization can effectively address the associated complex multiparameter optimization problem. The proposed numerical optimization framework can be extended to include additional parameters, such as atom number, interaction strength, imaging parameters, and experimental cycle time. The framework may therefore provide guidance for experimental implementation and for machine-learning-assisted quantum precision measurements with ultracold atoms.

\begin{acknowledgments}
This work is supported by the Key Special Project of the National Key Research and Development Program of China (Grant No. 2025YFF0515201), the Joint Fund for Quantum Major Research Plan of the National Natural Science Foundation of China (Grant No. U25D8014), the Artificial Intelligence Empowered Quantum Precision Measurement (Grant No. 40500-20101-222447), the Natural Science Foundation of Chongqing, China (Grant No. 2026NSCQ-GPX7586).

\end{acknowledgments}

\bibliography{ref}% Produces the bibliography via BibTeX.

\end{document}